\documentclass{article}
\usepackage{natbib}
\usepackage{emulateapj}

\bibpunct{(}{)}{;}{a}{}{,}

\newcommand{\SM}{$M_{\odot}$}

\begin{document}

\title{Metal Enrichment of The Primordial Interstellar Medium
through 3-D Hydrodynamical Evolution of The First Supernova Remnant}

\submitted{Appear in the Astrophysical Journal Letter}

\author{Naohito Nakasato and Toshikazu Shigeyama}

\affil{Research Center for the Early Universe,
Graduate School of Science, University of Tokyo, Bunkyo-ku, Tokyo 113-0033;
nakasato@astron.s.u-tokyo.ac.jp, shigeyama@astron.s.u-tokyo.ac.jp}

\begin{abstract}
The long-term evolution of supernova remnants (SNRs)
in the primordial interstellar medium (ISM)
with an inhomogeneous structure is calculated 
to investigate metal enrichment of the primordial gas.
For this purpose, we have constructed a parallel 3-D hydrodynamics code
incorporating the radiative cooling and self-gravity.
The self-gravity and radiative cooling develop
the inhomogeneous structure of the ISM
from a small perturbation with a power-law spectrum.
The resultant density ranges from 0.5 cm$^{-3}$ to 10$^6$ cm $^{-3}$.
Calculations for a supernova (SN) with the progenitor mass of 20 \SM\,
are performed as the first step of a series of our study.
It is found from the results that a single SN distributes
some of newly synthesized heavy elements into a dense
filament of the ISM with densities ranging from 100 to 10$^4$ cm$^{-3}$
depending on where the SN explodes.
Thus,  the metallicity [Mg/H] of the dense filaments polluted by the SN  
ejecta becomes  $-2.7\pm0.5$.
From these filaments, the first Population II stars will form.
This value is in accordance with the previous analytical work
(Shigeyama \& Tsujimoto) with accuracy of $\sim 0.3$ dex.
\end{abstract}


\section{Introduction}
Recently, the diversity in the abundance patterns of metal-deficient stars
in the Galactic halo has been observed by several authors
\citep{McWilliam_1995, Ryan_1996}.
On the other hand, theoretical attempts \citep{Audouze_1995, Shigeyama_1998}
have led to the conclusion that star formation induced
by quite few or a single supernova (SN) event
results in the observed diversity in the abundance patterns.
In particular, \citet{Shigeyama_1998} have suggested that
star formation occurs in the thin shell produced by a SN explosion
and that stars of the next generation inherit the abundance pattern thereof.
This shell was assumed to contain all the heavy elements newly synthesized
and ejected by the SN.
As a result, the abundance pattern of new stars is determined
by combinations of the SN ejecta and
the interstellar medium (ISM) swept up by the explosion.
In this scenario, the Rayleigh-Taylor instability is supposed
to happen to destroy the dense heavy element layer into numbers of blobs
which penetrate into the supernova remnant (SNR) shell.
It is clear that a spherically symmetric SNR in a uniform ISM cannot bring
its heavy element layer into the thin shell formed immediately behind
the shock front.
To check whether the abundance in the thin shell produced
by a SNR reaches the average abundance inside the SNR
as predicted by \citet{Shigeyama_1998}, it is necessary to calculate
hydrodynamical evolution of the SNR in three dimensions. 

The evolution of a SNR is characterized by several phases
\citep{Chevalier_1977} as the ejecta-dominated (ED) phase,
Sedov-Taylor (ST) phase, pressure-driven snowplow (PDS) phase, 
and momentum-conserving snowplow (MCS) phase.
Each phase has been investigated by a number of authors, e.g.,
for the first two phases see \citet{Truelove_1999}
and for the last two phases see \citet{Cioffi_1988}.
Most of these previous studies have been restricted
to spherically symmetric SNRs in the uniform ISM.
However, the inhomogeneity of an ISM affects the evolution of a SNR.
A model of SNR evolution in an ISM with a density gradient \citep{Hnatyk_1999}
has shown that the shape of the SNR was easily deformed to be non-spherical
in $\sim 10,000$ yr after explosion.
However, our objective is to see how a single SN event
distributes newly synthesized heavy elements in the ISM
when the SNR stops expanding at $t > 1$ Myr after explosion.
The ISM must also evolve on this time-scale
due to gravity and radiative cooling. 
Hence, we need to follow the evolution of a SNR
in the inhomogeneous and evolving ISM for more than 1 Myr.

In this Letter, we investigate the evolution of a SNR originated
from a Population III (Pop III) star in the inhomogeneous ISM
with the primordial abundances by using a 3-D hydrodynamics code.
Especially, we will focus our attention on the correlation of
the abundance of heavy elements with the gas density in each cell
to identify the abundance of heavy elements that would be inherited
by stars of the next generation.
Consequently, we can test the hypothesis proposed by \citet{Shigeyama_1998}.

\section{The method and initial model}
To model the three-dimensional evolution of a SNR, 
we have constructed a parallel 3-D hydrodynamics code
incorporating self-gravity and radiative cooling of
the ISM with no heavy elements.
We adopt the Godunov type \citep{Godunov_1959} scheme
to solve the hydrodynamical equations 
\citep[the HLLC method in][]{Toro_1997}.
The usage of the first order scheme reduces not only
the computational costs of the 3-D calculation
by a factor of two compared with higher order schemes like
PPM \citep{Colella_1984} but also the number of boundary values to be
transferred in parallel version.
To obtain a second order scheme in time, 
the Strang-type dimensional splitting \citep{Strang_1968} is used.
In addition, we have made two modifications to the usual Godunov type scheme.
First, to properly follow the adiabatic change of thermal energy
in a pre-shock region, we adopt the method used in a usual
cosmological hydrodynamics code \citep{Ryu_1993}.
Second, the consistent multi-fluid advection method \citep{Plewa_1999}
is implemented since we are interested in how heavy elements
ejected by a SN explosion mix with the ISM.
The 3-D Poisson equation is solved by the FFT method
to calculate the self-gravity of the ISM.
To solve the energy equation including radiative cooling term, 
we use an implicit method, i.e., the Newton-Raphson method.

In the early stage of the universe, the gas contains no heavy element and
the main coolant at temperatures below $10^4$ K is hydrogen molecules.
However, to solve rate equations for the formation and destruction of
hydrogen molecules is not feasible for a simulation like presented here.
We note that there are some attempts to take this approach
\citep[e.g.,][]{Abel_1998}.
Accordingly, throughout this Letter,
we use the radiative cooling rates for the gas with [Fe/H] $= -2$
\footnote[1]{The abundance ratios of the other heavy elements with respect to
Fe have the ``primordial'' values in \citet{Sutherland_Dopita_1993}}
to mimic the cooling function of the very early stage of the universe.
The cooling rate is computed by MAPPINGS III software
by R.S. Sutherland \citep[MAPPINGS III is the successor of MAPPINGS II described in][]{Sutherland_Dopita_1993}.
The radiative cooling due to hydrogen molecules is not
included in MAPPINGS III.
In the low temperature range as low as 1000 K, 
the adopted cooling rate is slightly lower than the cooling rate
by hydrogen molecules \citep[see Figure 2 of][]{Nakasato_2000a}.
We use the equation of states incorporating
hydrogen and helium in ionization equilibrium.
We have assumed that ions and electrons are in thermal equilibrium,
since the temperature in the SNR is almost always below $10^5$ K
during the evolution. 

The ISM is modeled by a periodic square box composed of 150$^3$ zones.
The size of the simulation box is set to 80 pc
to cover the maximum size of a SNR.
The simulation procedure is divided into following two stages.

The first stage of the simulation is devoted
to archiving an inhomogeneous ISM structure.
Firstly, an initial density field ($\rho(\vec{r})$)
for the first stage of the simulation is generated
by the COSMICS package \citep{Bertschinger_1995}
with a power spectrum of the density expressed as $P(k) \propto k^{-1}$.
The initial mean number density of hydrogen atoms is set to
$n = 100$ cm$^{-3}$.
We assume that the initial temperature is uniform with $T = 100$ K. 
Then, we obtain the initial velocity field ($v_i(\vec{r})$)
by integrating the equation of motion for one dynamical time
($t_{\rm dyn} \sim \sqrt{1/\left(4 \pi G \rho\right)} \sim 2.7$ Myr
in the present case, where $G$ is the gravitational constant).
We use the Zel'dovich approximation \citep{Zeldovich_1970} in this
integration: $v_{\rm i}(\vec{r}) =  a_{\rm i}(\vec{r}) t_{\rm dyn}$, where
$a_{\rm i}(\vec{r})$ is the acceleration field generated
by $\rho(\vec{r})$ through the Poisson equation.
Then we have followed the hydrodynamical evolution
of the box for further $5 t_{\rm dyn}$.
Due to the self-gravity, filamentary and knotty structures gradually form
like in calculations of the cosmological structure formation.
In $5 t_{\rm dyn}$ of the evolution, the inhomogeneity develops
in the box with the density ranging
from 0.5 cm$^{-3}$ to 10$^6$ cm$^{-3}$.
The volume filling factor of the low density region with $n < 10^{2}$ cm$^{-3}$
(this equals to the initial mean density) is $\sim 84$ \%.
The velocity dispersion is about 3 km s$^{-1}$.
This value will be used to test the hypothesis of \citet{Shigeyama_1998}. 
In the subsequent paper, we will present a detailed analysis of
the 3-D inhomogeneous structure of the ISM obtained
in the first stage of our simulation.

In the second stage of our simulation, we model the ST and PDS phases
of the evolution of a SNR with the same code.
For the initial condition of the SNR model, we deposit the SN explosion
energy ($E_{\rm sn} = 10^{51}$ erg) into a cell as thermal energy.
At the same time, we add the progenitor mass (20 \SM\, as a fiducial value)
to the same cell and treat 1.7 \% (0.34 \SM) of the progenitor mass as
Mg \citep[][]{Tsujimoto_1995}.

In this Letter, we {\it do not intend} to simulate
the star formation processes.
Instead, we show the results for two cases : 
the first SNR event occurs in (1) low density region or
(2) high density region.
We consider the region where $n$ is lower than $10^{2}$ cm$^{-3}$
as the low density region and the rest of the box as the high density region.
In the case (1), we select an arbitrary cell
in the low density region as a SN site.
After the star formation, the star leaves the star forming site and moves
around under the gravitational field of the ISM.
Thus, most of SNR events are expected to occur
in the low density regions due to the large volume filling factor.
Also in the case (2), we select an arbitrary cell
in the high density region as a SN site.
In the case (2), we treat 20 \SM\, of the gas in the cell
as the progenitor mass.

\section{Results}

\subsection{case (1): low density regions}
The first SN occurs in a smooth region where
the density gradient of the ISM is small.
This region corresponds to a ``hole'' (or a tenuous region) of
the inhomogeneous ISM.
The number density ($n$) of the chosen SN site
at the end of the first stage is $\sim 13$ cm$^{-3}$.
After adding 20 \SM\, ejecta to the cell,
$n$ becomes $\sim 5350$ cm$^{-3}$.
The temperature ($T$) of the SN site becomes $\sim 1.2 \times 10^{8}$ K
at the beginning of the second stage.
Then a strongly shocked region is produced and
the density of the cell decreases very quickly.
Although the ejecta soon become nearly spherical in $< 0.1$ Myr,
they are deformed after the collision with the dense structure of the ISM.

\begin{figure}
\epsscale{0.5}
\plotone{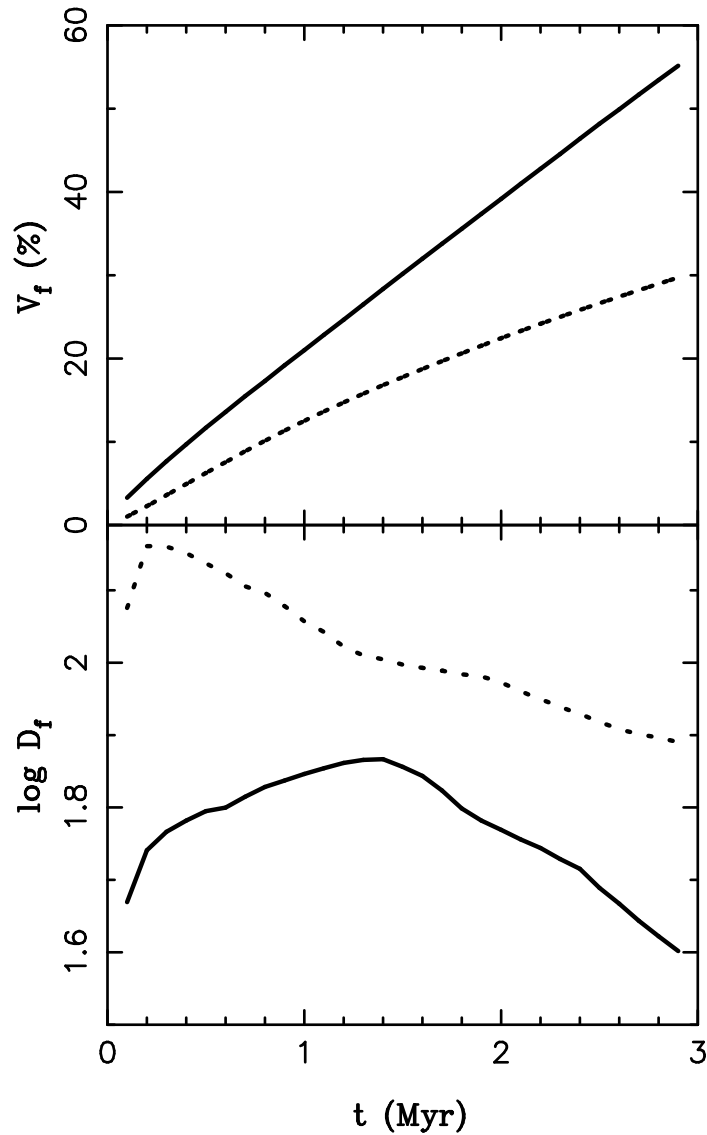}
\epsscale{1.0}
\caption{
Upper panel :
The evolution of the volume filling factor ($V_{f}$)
of the metal-enriched region.
Lower panel :
The evolution of the deformation factor ($D_{f}$)
of the metal-enriched region.
In both panel, the solid and dotted lines correspond
to the case (1) and (2), respectively.
\label{vol}
}
\end{figure}

To see the evolution of the shape of the ejecta quantitatively,
let $V$ the volume of the metal-enriched regions and $S$
the surface area of the metal-enriched regions.
We define the volume filling factor ($V_{f}$) as $V/L^3$ and
the deformation factor ($D_{f}$) as $S/(4\pi((3V)/(4\pi))^{2/3})$, 
where L is the size of the simulation box.
Figure \ref{vol} shows the evolutions of $V_{f}$ (upper panel)
and $D_{f}$ (lower panel) with the solid line for the case (1).
In the PDS phase of a spherical SNR in a uniform ISM,
the shocked volume evolve as $\propto t^{6/7}$ \citep{Cioffi_1988}.
If we assume the $V_{f}$ evolves as $\propto t^{\alpha}$, 
the exponent $\alpha$ during the first 1 Myr is smaller than unity
and then increases to almost unity at $t \sim 3$ Myr.
It is likely that the ejecta expand faster than the shock front
in later phases.
The fact that $D_{f}$ shown in the lower panel is
significantly greater than unity indicates that the shape of
the surface of the ejecta becomes very deformed
from a sphere due to the Rayleigh-Taylor instability. 

\begin{figure}
\plotone{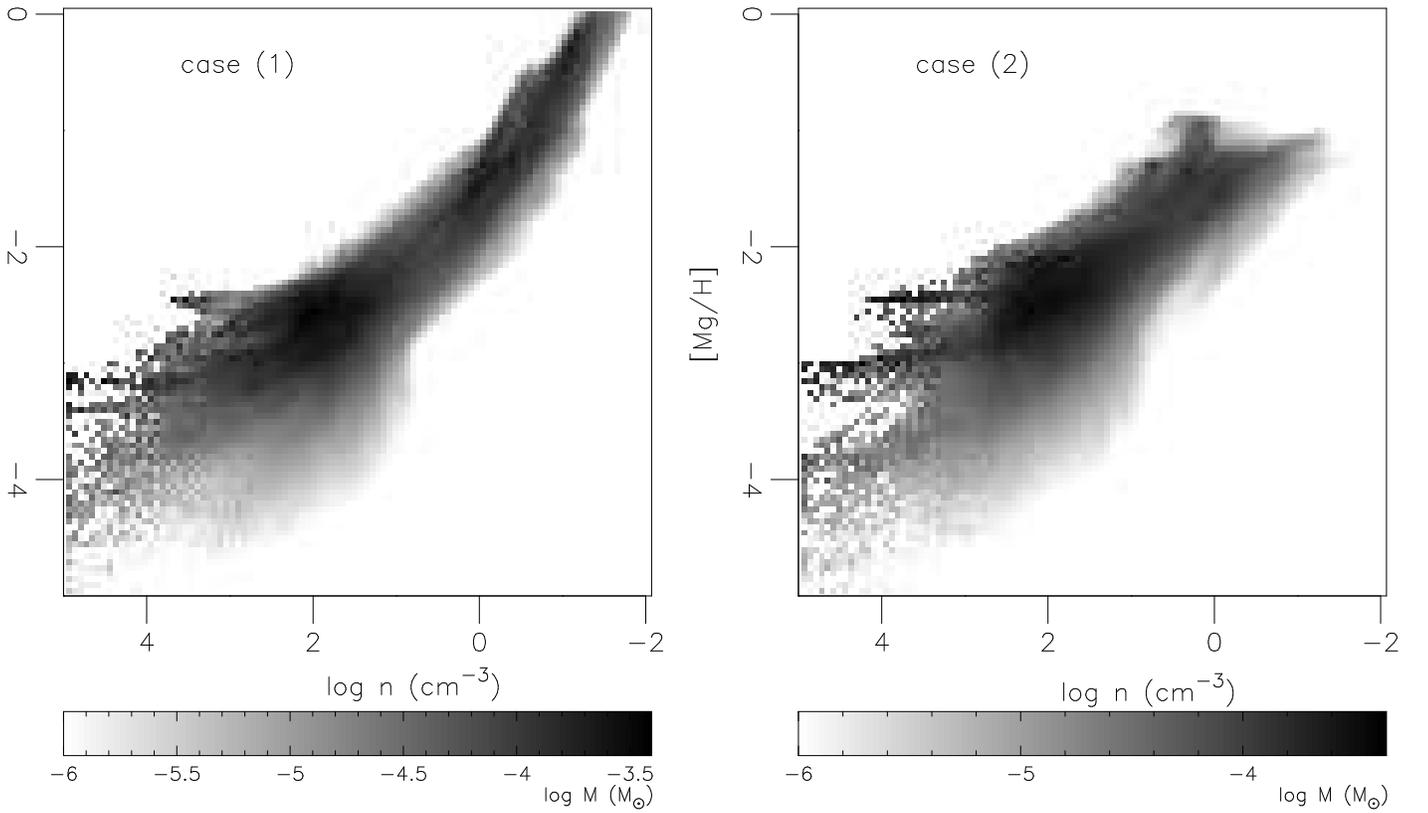}
\caption{
The mass of the heavy elements at $t = 3$ Myr
with number density ($\log n$) and given metallicity ([Mg/H]).
The left and right panels correspond to case (1) and case (2),
respectively.
\label{met}
}
\end{figure}

Figure \ref{met} shows the mass of the heavy elements at $t = 3$ Myr
as a function of number density ($\log n$) and metallicity ([Mg/H]).
We note that this mass distribution function does not change its
shape for last 1 Myr.
There are two peaks in this diagram.
One is located at $(\log n,\, {\rm [Mg/H]}) \sim (0,\,-2)$
and not so pronounced as the other.
The total amount of the heavy elements involved in this peak is not so large.
The low density indicates that these heavy elements have been staying 
near the SN site.
The other peak is located at $(\log n,\, {\rm [Mg/H]}) \sim (2.1,\,-2.7)$.
A majority of the heavy element is involved in this peak.
The swept up mass estimated from equation (2) of \citet{Shigeyama_1998}
becomes $\left(10^{2.1}\right)^{-0.062}\times\left(3/10\right)^{-9/7}\sim 3.5$
times greater than the original value in this equation.
Here, the first factor in the left hand side comes from the density
and the second from the sound speed (or the velocity dispersion). 
Thus the [Mg/H] estimated from this equation would be $\sim -2.9$.
The analytical work and 3-D calculation have shown
a fairly good agreement in this respect.
What \citet{Shigeyama_1998} did not mention is
that this peak has a finite width.
This metallicity distribution can be considered to be a consequence of the  
inhomogeneity of the ISM: A part of the ejecta that penetrates into a dense  
part of the primordial ISM, mixes with it, and tends to get a small  
metallicity. On the other hand, a part of the ejecta that contacts with a  
tenuous part of the ISM, mixes with a small amount of ISM thereof, and thus  
gets a larger metallicity. A part of the ejecta that never mixes with the  
ISM will keep its initial metallicity.

The resultant [Mg/H]  ranges $-3.2 <$ [Mg/H] $<-2.2$
with a peak at [Mg/H] $\sim -2.7$.
With the resolution of our calculations, i.e. $dx \sim 0.53$ pc,
a cell with $\log n > 2.43$ includes more than one solar mass.
Thus the metallicity inherited by the next generation stars may be
overestimated unless the mass of these stars is much smaller
than one solar mass.  

\subsection{case (2): high density regions}
\begin{figure}
\plotone{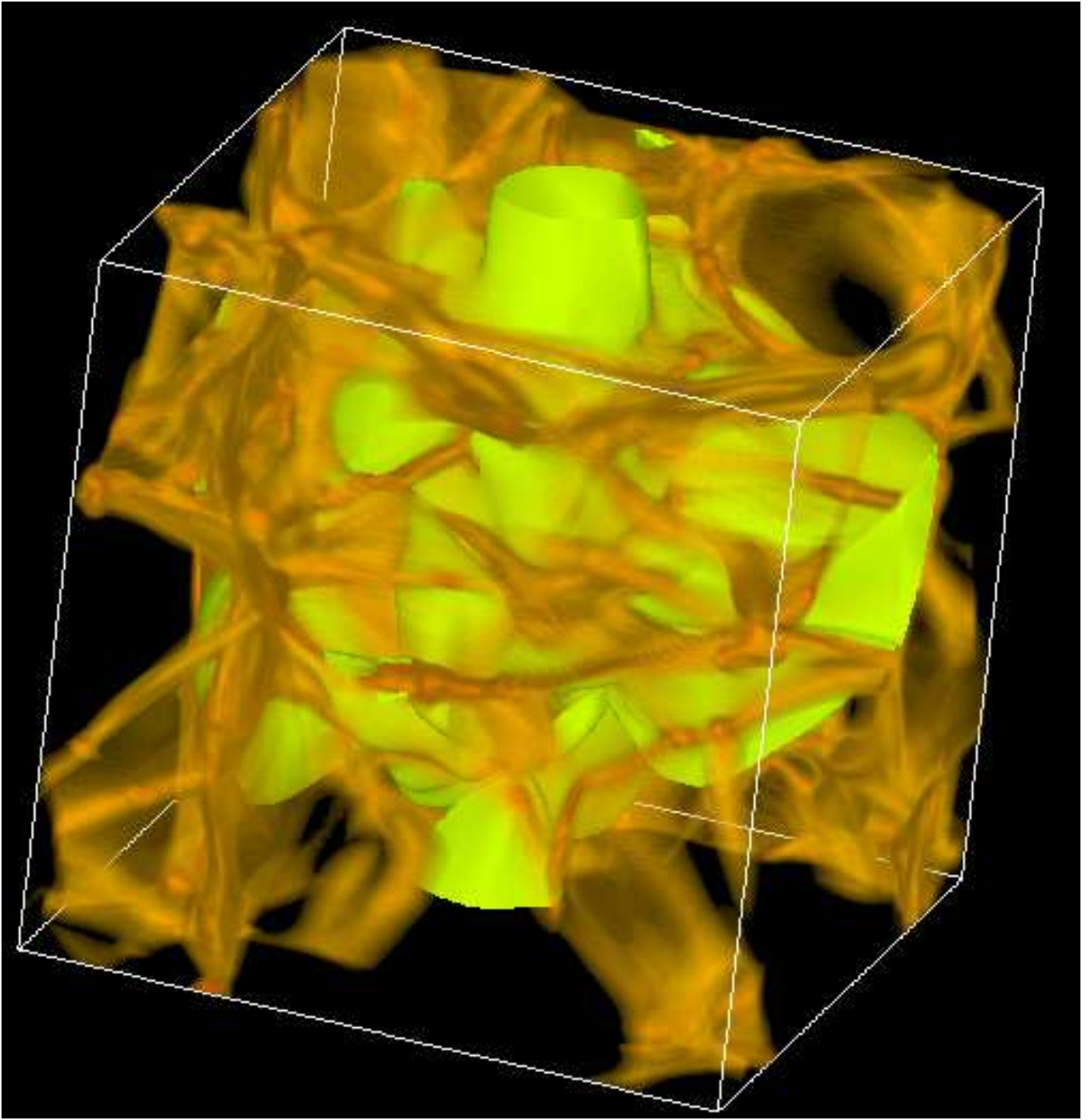}
\caption{
The metal-enriched region at $t = 3$ Myr
for case (2) is rendered by a yellow shaded iso-surface.
The surface value corresponds to [Mg/H] $\sim -2.6$.
The orange filamentary shape is a volume rendering of
the dense structure at the same time.
The SN site is located at the center of the simulation box
shown with the white lines. 
\label{3d}
}
\end{figure}

In this case, $n$ and $T$ of the SN site at the beginning of the second stage
are $\sim$ 11,500 cm$^{-3}$ and $5.6 \times 10^7$ K, respectively.
The evolution of $V_{f}$ and $D_{f}$ is presented in Figure \ref{vol}
with the dotted lines.
We can see that $V_{f}$ is always smaller than in case (1) and 
$D_{f}$ is always higher than in case (1).
This is because the SN occurs in a dense filament
that produces a large density gradient.
Namely, the SNR easily expands perpendicular to the direction
of the filament but is much slowed down along the direction of the filament.
The surface of the metal-enriched region at $t = 3$ Myr
is shown in Figure \ref{3d} together with the filamentary structure of the ISM.
The SN site is located at the center of the simulation box.
The yellow surface represents the [Mg/H] $= -2.6$ iso-surface 
and the orange filamentary shape shows the 3-D density structure.
The ejecta expand into tenuous regions of the ISM and are
deformed by the filamentary ISM structure.

As in case (1), the frequency distribution function of the mass
of Mg at $t = 3$ Myr with respect to $\log n$ and [Mg/H]
is shown in the right panel of Figure \ref{met}.
Although there are two peaks at $(\log n,\, {\rm [Mg/H]}) \sim(2.3,\,-2.6)$ and
$(\log n,\, {\rm [Mg/H]}) \sim (4.0,\,-2.5)$ also in this case,
both of the locations and shapes are different from those for case (1).
Equation (2) of \citet{Shigeyama_1998} gives [Mg/H] $\sim -2.8$ that
agrees with this 3-D calculation within a factor of $\sim 2$.

\section{Summary and Discussion}
In this Letter, we have presented the first results of 3-D
hydrodynamical calculations of a SNR originated from a Pop III star
in the inhomogeneous ISM with the primordial abundances.
We have constructed an efficient parallel numerical code to
solve the 3-D hydrodynamical equations and the Poisson equations.
Using our code, we first evolve a random density field and obtain
the inhomogeneous ISM model.
Then, we follow the long-term evolution of a SNR to see
how a single SN distributes the newly synthesized
heavy elements into the ISM.

For the site of the first SN event in the simulation,
we consider low density and high density environments.
In both cases, a strong shock is produced and then the shock front
is slowed down by collisions with the filamentary structure of the ISM.
Our calculations indicate that a majority of the newly synthesized
heavy elements is distributed into the gas in filamentary structures
to get the metallicity of $-3.2 \leq$ [Mg/H] $\leq -2.2$
depending on the SN site.
Thus, Pop II stars born from a high density filament
polluted with the ejecta of a SN will have [Mg/H] $\sim -2.7$
on their surfaces.
We have compared the values of [Mg/H] in dense regions of the simulation box
with those estimated from the analytical work by \citet{Shigeyama_1998} and
found that they are in agreement with each other with an accuracy
to $\sim 0.3$ dex, though the assumed density structure of the ISM
in \citet{Shigeyama_1998} is quite different from that in the present work.

To understand metal enrichment process at the early stages of
galaxy evolution, we will need to eliminate (or identify, at least)
the influences from numerical artifacts that may be caused
by the limited resolution of our model.
First, we needed to deposit thermal energy into one cell to initiate a SN.
A significant fraction of the explosion energy of a real SN
with the size of $\sim 0.5$ pc must be in the form of kinetic energy.
Thus the ejecta of a real SN are expected to penetrate deeper into the
filamentary structure to get lower metallicity than shown in this Letter.
Second,  the inhomogeneous ISM model obtained in the first stage of
our simulation procedure may be changed depending on a number of
model assumptions which include the initial power spectrum of the model,
the cooling rates due to molecular hydrogen,
neglect of the dark matter potential, the effect of radiative transfer,
and etc..

We will report results of a detailed study of the inhomogeneous ISM and  
longer evolution of the metallicity distribution in a
separate paper \citep{Nakasato_2000b}. Calculations for SNe with different
progenitor masses will be also performed to see the abundance patterns of  
Pop II stars.

\end{document}